\documentclass[conference]{IEEEtran}
\IEEEoverridecommandlockouts
\usepackage{cite}
\usepackage{amsmath,amssymb,amsfonts}
\usepackage{algorithmic}
\usepackage{graphicx}
\usepackage{textcomp}
\usepackage[table]{xcolor}
\usepackage{xcolor}
\usepackage{balance}
\usepackage{booktabs}

\def\BibTeX{{\rm B\kern-.05em{\sc i\kern-.025em b}\kern-.08em
    T\kern-.1667em\lower.7ex\hbox{E}\kern-.125emX}}
\begin{document}

\title{Smart Environmental Monitoring of Marine Pollution using Edge AI\\

\thanks{The authors thank the funding support of the Carl Zeiss Stiftung under the Sustainable Embedded AI project (P2021-02-009).}
}

\author{
\IEEEauthorblockN{Mohamed Moursi, Norbert Wehn, Bilal Hammoud}
\IEEEauthorblockA{\textit{Microelectronic Systems Design (EMS) RPTU, Kaiserslautern-Landau}, Germany\\ Email:\{mmoursi, norbert.wehn, bilal.hammoud\}@rptu.de
}
}

\maketitle

\begin{abstract}
Oil spill incidents pose severe threats to marine ecosystems and coastal environments, necessitating rapid detection and monitoring capabilities to mitigate environmental damage. 
In this paper, we demonstrate how artificial intelligence, despite the inherent high computational and memory requirements, can be efficiently integrated into marine pollution monitoring systems. 
More precisely, we propose a drone-based smart monitoring system leveraging a compressed deep learning U-Net architecture for oil spill detection and thickness estimation. Compared to the standard U-Net architecture, the number of convolution blocks and channels per block are modified. The new model is then trained on synthetic radar data to accurately predict thick oil slick thickness up to 10 mm. Results show that our optimized Tiny U-Net achieves superior performance with an Intersection over Union (IoU) metric of approximately 79\%, while simultaneously reducing the model size by a factor of $\sim$269x compared to the state-of-the-art.
This significant model compression enables efficient edge computing deployment on field-programmable gate array (FPGA) hardware integrated directly into the drone platform. Hardware implementation demonstrates near real-time thickness estimation capabilities with a run-time power consumption of approximately 2.2 watts. Our findings highlight the increasing potential of smart monitoring technologies and efficient edge computing for operational characterization in marine environments.
\end{abstract}


\section{Introduction}
Oil spills represent a significant environmental hazard with high impacts on marine ecosystems and wildlife populations, constituting one of the most severe anthropogenic disturbances to maritime environments. 
It is imperative from both economic and ecological perspectives to mitigate the deleterious effects of these spills \cite{ugwu2021ecological}.
While state-of-the-art research has predominantly focused on oil detection capabilities, a detailed characterization of additional spill parameters like the spatial distribution of oil slick thickness helps implement effective containment strategies \cite{fingas2018challenges}. 

The development of advanced monitoring systems that spontaneously provide such detailed characterization enables more precise intervention approaches and ultimately improves remediation outcomes. This requires careful consideration of the selected platform and sensor as potential candidates for specific monitoring functionalities \cite{fingas2017review}. Drone platforms have received attention for environmental monitoring in the last years \cite{zhang2023review}, considering their relatively low cost, high resolution, and fast intervention.

For oil spill monitoring, several drone-based approaches targeted oil detection using visual \cite{jiao2019new}, and infrared cameras \cite{de2020oil}. Hyperspectral images were also analyzed to provide thickness estimation \cite{jiang2021hyperspectral}. But the previous sensors cannot operate during the full daytime and in different weather conditions. For this, the radar sensor is more prominent for monitoring purposes, which explains why synthetic aperture radar (SAR) imaging have been extensively used for oil spill detection \cite{li2024novel,trujillo2024marine,yang2024near}. However, even when processing SAR imaging with advanced deep-learning models, the effectiveness of this technique is reduced in calm ocean conditions due to the loss in the radar backscattering \cite{skrunes2014comparing}. Additionally, SAR systems are rarely used for spatial thickness estimation. 

Most state-of-the-art drone-based solutions lack on-site real-time processing due to the constraints imposed by the limited power budget of battery-powered drones. To address this, field-programmable gate array (FPGA) compute platforms have been widely explored in other applications \cite{di2024board,briley2024hardware} for executing complex algorithms at the edge. However, their use in oil spill monitoring remains largely unexplored.

Contrary to the state-of-the-art techniques, we suggested a drone platform with radar sensor approaches for both, oil spill detection \cite{hammoud2023unet} and thickness estimation \cite{hammoud2022multidimensional}, in calm ocean conditions. Additionally, we proposed extremely small deep learning models (few bytes in size) that are suitable for direct onboard processing \cite{hammoud2024fpga}. But these TinyML models work only at extremely low wind speeds. Then, we improved the monitoring by covering moderate wind speed scenarios using a more complex U-Net model \cite{hammoud2024date}.

In this paper, we propose a drone-based smart monitoring system using an optimized Tiny U-Net model for on-site estimation of up to 10 mm oil slick thickness at both calm and moderate wind speeds. This work differs from that of \cite{hammoud2024date} by optimizing the architecture of the original U-Net to further boost the system and model performance.
By applying different compression techniques, the optimized Tiny U-Net outperforms the state-of-the-art in estimation performance, while reducing its size by a factor of ~269x. Furthermore, by implementing the Tiny U-Net on the FPGA with memory constraints, the applied optimization showcases the efficient edge computing capability of the proposed system compared to \cite{hammoud2024date} in near real-time with low power consumption. 

\begin{figure*}[h]
	\begin{center}
		\includegraphics[width=2\columnwidth]{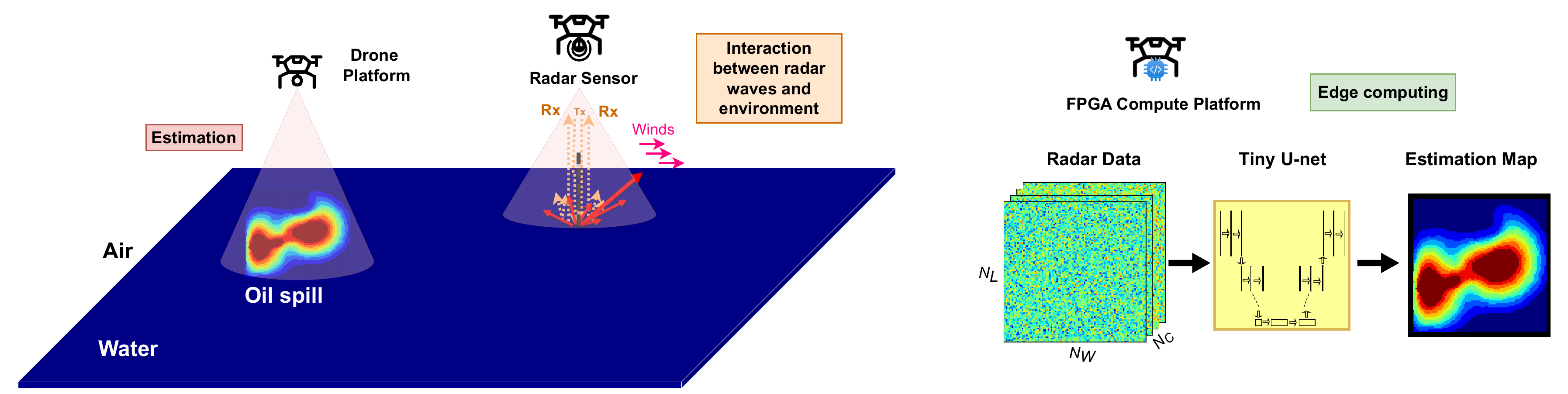}
		\caption{The drone-based proposed system for oil spill monitoring with on-edge computing.}  
		\label{fig:fig1}
	\end{center}
\end{figure*}

\section{Monitoring System}

\subsection{Environmental Model}
For the marine environment, we follow the same model adopted in \cite{hammoud2024date}. The current investigation focuses on floating thick slicks that occur at the early stage (within hours to 1-2 days) of the spill before emulsification, with a thickness range from 1 to 10 mm. The ocean surface roughness is mathematically modeled by the surface root-mean-square height. It depends on the wind speeds above the sea surface, which is varied between 2 and 8 m/s. The relative dielectric constant of the oil is assumed to be 3. The water temperature and salinity are assumed 20$^\circ$ and 35 ppt, respectively.

\subsection{Monitoring System Model}

The target is to estimate the spatial thickness distribution of the oil spill directly on-site. Our system is composed of three main components, as shown in Fig.~\ref{fig:fig1}: 
\begin{itemize}
    \item The drone platform to carry the sensor and perform the scanning of potential oil spill scenes.
    \item A wide band radar sensor that utilizes multiple frequency channels in C and X -bands (4-12 GHz), which are convenient considering the resulting antenna size and weight relative to the drone. The selected number of frequencies is 9, from 4 GHz to 12 GHz with an increment of 1 GHz.
    \item An FPGA hardware compute platform as an embedded system mounted on the drone platform to perform on-edge data processing.
\end{itemize}
 
During the scan of a potential oil spill scene, the drone-based radar operates as a nadir-looking system where the propagating electromagnetic waves are normally incident on the ocean surface. Thus, the radar specular backscattering is completely captured \cite{hammoud2019experimental}. Afterwards, the radar backscattering is processed onboard on the FPGA.
The synthetic radar dataset is based on the complete framework presented in \cite{hammoud2024date}, and publicly available in \cite{boumaroun2024github}.

\section{Tiny U-Net Model}

\begin{figure*}
	\begin{center}
		\includegraphics[width=1.8\columnwidth]{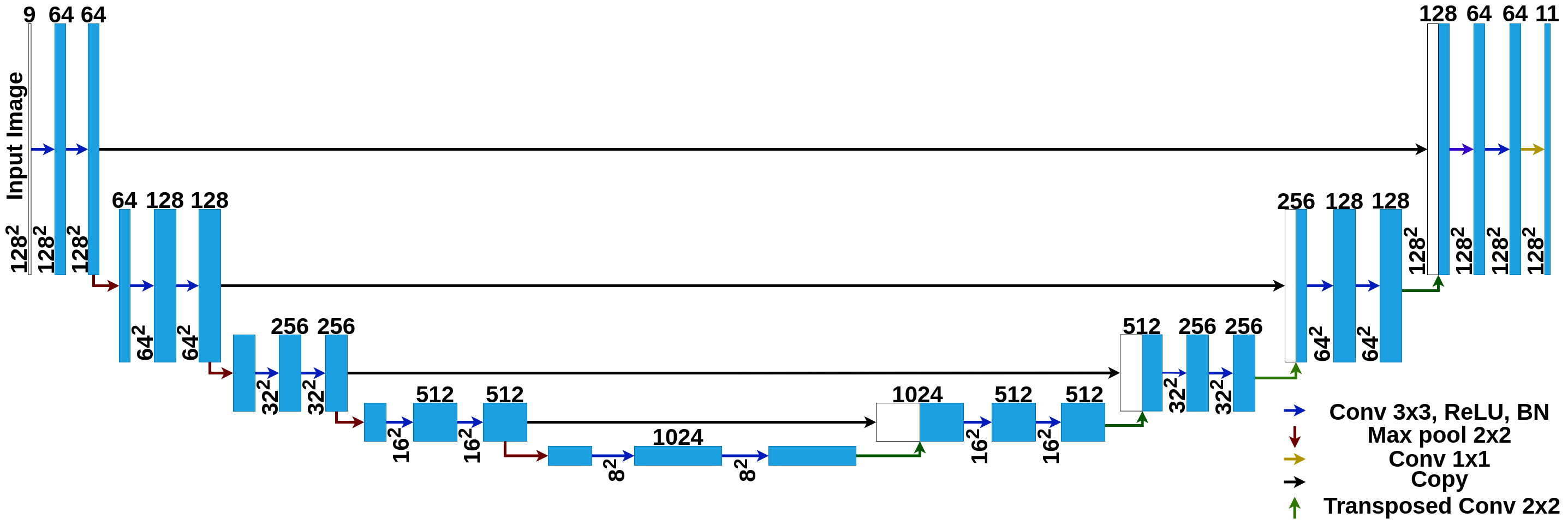}
		\caption{U-Net internal architecture.}  
		\label{fig:fig2}
	\end{center}
\end{figure*}

\subsection{Input and Output Maps}
Once the wideband radar scans the scene, the collected data is forwarded to the U-Net model in a 3-D $N_w \times N_l \times N_c$ matrix. $N_w$ and $N_l$ represent respectively the geometrical width and length of the scanned environment, and $N_c$ represents the number of used radar channels.

The model outputs the spatial distribution map of the estimated thickness of potential oil slicks.
Considering the discrete range of thicknesses is from 0 mm to 10 mm, each pixel in this output map is assigned to one of 11 thickness classes. The class 0 mm corresponds to the detection of clean water surface scenarios rather than the estimation.

\subsection{Architecture Design Space Exploration}
As shown in Fig.~\ref{fig:fig2}, the original architecture used in \cite{ronneberger2015u} is composed of 8 convolution blocks. Each block is composed of two convolutional layers followed by batch normalization layers and rectified linear unit (ReLU) activation functions. The downward pass, also known as the encoder, is composed of 4 convolution blocks followed by a max pooling layer each. The other 4 blocks, which are followed by transpose convolution layers, are responsible for the upward pass,
also known as the decoder. Finally, the layers in the encoder part are skip-connected and concatenated with layers in the decoder part. More details regarding the U-Net model architecture can be found in \cite{ronneberger2015u}.

Considering that our goal is to make it feasible to run the U-Net model efficiently on edge, we explore in this work two dimensions of the model architecture to reduce the network size:
\begin{enumerate}
    \item the number of convolution blocks, denoted by $B$, for the encoder or decoder in the model architecture. $B$ can have a minimum value of 1 and a maximum value of 4. 
    \item the number of channels for each convolutional layer in the convolution block, which is reduced by a factor ~ $F$~$=1,2,$~etc.
\end{enumerate}
The aim of reducing these dimensions is to compress the original large U-Net to a Tiny U-Net model, which fits in size on an FPGA with limited memory, while not affecting the overall estimation performance. 

\subsection{Training Process}
To explore the design space of different potential architectures of the U-Net model, we use 555 oil spill scenarios from \cite{boumaroun2024github} for testing, 504 of which are used for training whereas the remaining 51 are used for validation. To train the models, the categorical cross-entropy loss function is used. During the training process, all inputs are normalized before being processed by the U-Net architectures, making their mean and standard deviation 0 and 1, respectively. This helps in reaching convergence during the training process. The models are trained on 10 epochs with a batch size of 16 and a learning rate of 0.0008. The training process is performed using the PyTorch Python library for 32-bit floating-point simulations. 
For efficient hardware implementation, the model parameters have been quantized to 8 bits using the Open Neural Network Exchange (ONNX) post-training quantization.

\subsection{Evaluation Metrics}
We use the following four metrics to assess the estimation performance of different compressed architectures: the intersection over union (IoU), precision, dice, and recall, as defined below. Since our problem is a multi-class segmentation, with $N =$ 11 classes for different oil thicknesses ranging from 0 to 10 mm, we average the metrics taken from each class.

\begin{equation}\label{eq:mean_iou}
	\text{Mean IoU} = \frac{1}{N}\sum_{n=1}^{N} \frac{\text{TP}_{n}}{\text{TP}_{n} + \text{FP}_{n} + \text{FN}_{n}},
\end{equation}

\begin{equation}\label{eq:mean_precision}
	\text{Mean Precision} = \frac{1}{N}\sum_{n=1}^{N} \frac{\text{TP}_{n}}{\text{TP}_{n} + \text{FP}_{n}},
\end{equation}

\begin{equation}\label{eq:mean_recall}
	\text{Mean Recall} = \frac{1}{N}\sum_{n=1}^{N} \frac{\text{TP}_{n}}{\text{TP}_{n} + \text{FN}_{n}},
\end{equation}

\begin{equation}\label{eq:mean_dice}
	\text{Mean Dice} = \frac{1}{N}\sum_{n=1}^{N} \frac{2 . \text{TP}_{n}}{ (2 . \text{TP}_{n} + \text{FP}_{n}+ \text{FN}_{n})}.
\end{equation}

TP, TN, FP, and FN represent, respectively, the true positive, true negative, false positive, and false negative pixels in the output matrix of the model.

\section{Results}

\begin{table*}[t]
    \centering
    \caption{Comparison of estimation performance and model size for different combinations of dimensions reduction}
    \label{tab:comparison}
    \renewcommand{\arraystretch}{1.2}
    \begin{tabular}{cc|cccc|cccc|c}
        \hline
        \textbf{B} & \textbf{F} & \multicolumn{4}{c|}{\textbf{Floating Point}} & \multicolumn{4}{c|}{\textbf{Quantized}} & \textbf{Model Size (MB)} \\
        & & IoU & Dice & Precision & Recall & IoU & Dice & Precision & Recall & \\
        \hline
        \rowcolor{red!20} 4 & 1  & 0.70 & 0.80 & 0.79 & 0.84 & 0.70 & 0.80 & 0.79 & 0.85 & 29.62 \\
        4 & 2  & 0.76 & 0.86 & 0.86 & 0.86 & 0.76 & 0.86 & 0.86 & 0.86 & 7.41  \\
        4 & 4  & 0.75 & 0.84 & 0.85 & 0.84 & 0.75 & 0.84 & 0.85 & 0.84 & 1.86  \\
        4 & 8  & 0.59 & 0.71 & 0.71 & 0.73 & 0.62 & 0.73 & 0.73 & 0.73 & 0.47  \\
        4 & 16 & 0.09 & 0.13 & 0.14 & 0.15 & 0.08 & 0.11 & 0.10 & 0.12 & 0.12  \\
        \hline
        3 & 1  & 0.78 & 0.86 & 0.87 & 0.86 & 0.78 & 0.86 & 0.87 & 0.86 & 7.35  \\
        3 & 2  & 0.77 & 0.85 & 0.85 & 0.87 & 0.77 & 0.85 & 0.85 & 0.86 & 1.84  \\
        3 & 4  & 0.75 & 0.84 & 0.84 & 0.85 & 0.75 & 0.84 & 0.84 & 0.85 & 0.46  \\
        3 & 8  & 0.66 & 0.75 & 0.73 & 0.77 & 0.66 & 0.74 & 0.73 & 0.76 & 0.12  \\
        3 & 16 & 0.19 & 0.24 & 0.26 & 0.29 & 0.20 & 0.25 & 0.25 & 0.28 & 0.03  \\
        \hline
        2 & 1  & 0.78 & 0.86 & 0.87 & 0.86 & 0.78 & 0.86 & 0.87 & 0.85 & 1.78  \\
        2 & 2  & 0.78 & 0.86 & 0.86 & 0.86 & 0.78 & 0.86 & 0.86 & 0.86 & 0.45  \\
        \rowcolor{green!30} 
        2 & 4  & 0.79 & 0.85 & 0.85 & 0.87 & 0.79 & 0.86 & 0.85 & 0.87 & 0.11  \\
        2 & 8  & 0.63 & 0.72 & 0.75 & 0.70 & 0.62 & 0.70 & 0.74 & 0.69 & 0.03  \\
        2 & 16 & 0.16 & 0.21 & 0.22 & 0.26 & 0.19 & 0.24 & 0.28 & 0.27 & 0.01  \\
        \hline
        1 & 1  & 0.75 & 0.85 & 0.86 & 0.85 & 0.75 & 0.85 & 0.86 & 0.84 & 0.39  \\
        1 & 2  & 0.75 & 0.85 & 0.86 & 0.85 & 0.75 & 0.85 & 0.86 & 0.85 & 0.10  \\
        1 & 4  & 0.70 & 0.77 & 0.76 & 0.80 & 0.68 & 0.76 & 0.75 & 0.78 & 0.03  \\
        1 & 8  & 0.26 & 0.37 & 0.33 & 0.44 & 0.27 & 0.37 & 0.34 & 0.44 & 0.01  \\
        1 & 16 & 0.24 & 0.36 & 0.37 & 0.42 & 0.24 & 0.34 & 0.36 & 0.37 & 0.002  \\
        \hline
    \end{tabular}
\end{table*}

\subsection{Optimized Architecture}
Table~\ref{tab:comparison} presents the estimation performance results using both floating-point and quantized value simulations, for different combinations of compression parameters $B$ and $F$. For example, $B=4$ means that we have 4 convolution blocks in the encoder and another 4 in the decoder. Also, $F=2$ implies that the number of channels per convolutional layer is reduced to half. 
The first row in Table~\ref{tab:comparison} corresponds to the original U-Net architecture, whereas the last row corresponds to the maximal compressed model, as clearly shown by the model size.

According to the results obtained, the optimized U-Net in terms of estimation performance is highlighted in green. It has $B=2$, $F=4$, and the achieved IoU is the highest among all other possible architectures. Therefore, the optimized Tiny U-Net outperforms the original model by 9\% (highlighted in red), while its size is reduced by 269x.
It is worth noting here that for the calculation of the model size, we are omitting the required memory for the skip connections for all combinations.

\subsection{Performance Analysis}
Table~\ref{table:results_comparison} compares the estimation performance of the suggested Tiny U-Net to other state-of-the-art deep learning techniques, including the artificial neural network (ANN) \cite{hammoud2022artificial} and the two-level nested U-structure (U2-Net) architecture \cite{qin2020u2} that is composed of multiple U-Net
networks. 
The Tiny U-Net segmentation model outperforms all
models in all metrics, which indicates that this model is highly reliable in the correct estimation of each thickness class among others.
This demonstrates a significant advancement over existing methodologies by enabling accurate estimation of oil slick thickness up to 10 mm, thereby addressing a critical operational gap in marine pollution monitoring.

\begin{table}
\caption{Comparison of estimation performance to State-of-the-art\label{table:results_comparison}}
\centering
\renewcommand{\arraystretch}{1.2}
\begin{tabular}{ccccc} 
\hline 
\textbf{Model}	& \textbf{IoU}	& \textbf{Dice} & \textbf{Precision} & \textbf{Recall}\\
\hline 
ANN	\cite{hammoud2022artificial}	&  0.36		    &  0.49 	& 0.49 	    & 0.56\\
U2-Net \cite{qin2020u2}		        &  0.64			& 0.74 	    & 0.74 	    & 0.74\\
U-Net \cite{hammoud2024date}	& {0.75} & {0.86} & {0.85}  & {0.87}\\
\textbf{Tiny U-Net}	& \textbf{0.79} & \textbf{0.86} &\textbf{0.85}  & \textbf{0.87}\\
\hline
\end{tabular}
\end{table}

\subsection{Efficient Edge Computing}
Since our approach involves deploying drone-mounted radars for tactical response, it is essential to evaluate the feasibility of implementing the Tiny U-Net architecture on real hardware for onboard processing. To achieve this, we selected the Zynq UltraScale+ ZCU104 FPGA as the drone's compute platform and optimized the hardware design to enable efficient edge inference. 
Table~\ref{tab:utilization} shows the hardware utilization and evaluation metrics for the implementation of the Tiny U-Net architecture. The measured IoU on board shows no degradation in the estimation performance compared to simulation results. Additionally, the applied compression techniques improved the efficiency of the model implementation. I.e, results show that the Tiny U-Net fits within the FPGA constraints, where it uses 56\% of BRAMs after being reduced 269x from the original architecture. Without the compression, the implementation of the original U-Net would not be possible on this platform. In addition to the size reduction, the hardware implementation on the FPGA achieved milliseconds processing latency while maintaining low power consumption (approximately 2.2 watts), thus confirming the viability of deploying complex deep learning models on power-constrained aerial platforms.

\begin{table}
    \centering
    \caption{Hardware Evaluation Metrics}
    \label{tab:utilization}
    \renewcommand{\arraystretch}{1.2}
    \begin{tabular}{lcc}
        \toprule
        \textbf{Resources} & \textbf{Utilization} & \textbf{Percentage} \\
        \midrule
        LUT     & 30,094  & 13.06\% \\
        LUTRAM  & 3,999   & 3.93\% \\
        FF      & 27,328  & 5.93\% \\
        BRAM    & 175     & 56.09\% \\
        DSP     & 212     & 12.27\% \\
        \midrule
        \midrule
        \textbf{Performance} & \textbf{Tiny U-Net} & \textbf{U-Net \cite{hammoud2024date}} \\
        \midrule
        Power (Run-time) & 2.2 W & $-$\\
        Dynamic Power & \textbf{203 mW} & $\sim$ 400 mW \\
        Latency & \textbf{27.5 ms} & 2.2 s\\
        \midrule
        \textbf{IoU} on FPGA & \textbf{0.79} & $-$\\
        \bottomrule
    \end{tabular}
\end{table}

\section{Conclusion}
In this work, we proposed a drone-based smart monitoring system that integrates a compressed deep learning U-Net architecture for oil spill detection and thickness estimation. By optimizing the network structure through modifications in the number of convolution blocks and channels per block, our model effectively predicts thick oil slicks up to 10 mm using synthetic radar data. Results demonstrated that our optimized Tiny U-Net maintains a high estimation performance with an IoU of approximately 79\%, while achieving a substantial 269x reduction in model size compared to state-of-the-art approaches.
To enable real-time deployment, we implemented the model on an FPGA-based edge computing platform. The hardware design achieved near real-time thickness estimation with an inference latency of milliseconds and a low run-time power consumption of 2.2 watts, without any degradation in estimation performance. These results confirm the feasibility of deploying AI-driven oil spill monitoring systems directly on autonomous aerial platforms, enabling rapid and efficient environmental assessment. Future work will explore further optimizations to the model architecture and test our approach on a real dataset.

\balance
\bibliographystyle{IEEEbib}
\bibliography{references}

\end{document}